\documentclass[fleqn,10pt]{wlscirep}
\usepackage[utf8]{inputenc}
\usepackage[T1]{fontenc}
\usepackage{iopams}
\usepackage{amssymb}
\usepackage{amsfonts}
\usepackage{revsymb}
\usepackage{graphicx}
\usepackage{color}
\usepackage{xcolor}
\usepackage{soul}
\usepackage{caption}
\usepackage{subcaption}
\usepackage{siunitx}
\expandafter\let\csname equation*\endcsname\relax
\expandafter\let\csname endequation*\endcsname\relax
\usepackage{amsmath}
\usepackage{comment}

\providecommand{\U}[1]{\protect\rule{.1in}{.1in}}

\begin{document}

\title{Influence of Physical Properties of Hockey Stick Blade on Shots}

\author[1, 2]{Martin Plesch}
\affil[1]{Institute of Physics, Slovak Academy of Sciences, Bratislava, Slovak Republic}
\affil[2]{Institute of Computer Science, Masaryk University, Brno, Czech Republic}

\author[3]{Samuel J\'an Plesn\'ik}
\affil[3]{Prv\'e S\'ukromn\'e Gymn\'azium, Bratislava, Slovak Republic}

\author[4, *]{Nat\'alia Ru\v zi\v ckov\'a}
\affil[4]{Institute of Science and Technology, Klosterneuburg, Austria}
\affil[*]{natalia.ruzickova@ist.ac.at}

\begin{abstract}

Parameters of a shot of an ice hockey player are mostly determined by the capabilities of the player and the physical properties of the stick used. To reach better performance, every hockey player uses also a hockey tape for an adjustment of the stick blade, that changes both the damping properties as well as the friction coefficient of the blade surface. To show the unexpected extent to which these physical properties of the blade affect the shot, we compared two types of blade cover: traditional tape (rolled onto the blade) and a blade sticker that adheres to both sides of the hockey stick blade. We analysed high-speed recordings of two types of shots by 13 players, ranging from amateurs and junior players to NHL superstars. The two covers differ greatly in friction coefficient and stiffness, which results in significantly (more than $99\%$ confidence) greater speed, rotation and energy when using the stiffer and rougher sticker.   

\end{abstract}

\flushbottom
\maketitle

\thispagestyle{empty}


\section{Introduction}
Ice hockey sticks have experienced a gradual development in recent decades. Progress in material science allowed the replacement of widely used wood by composite materials that nowadays represent the standard for hockey players worldwide. 

Broad usage of new materials like aluminium and composites has become an interesting topic for research as well. Interestingly, the performance of wooden sticks was shown to be superior to the performance of composite sticks \cite{bigford}. Stick stiffness has been proven to directly influence performance \cite{kays} and the hockey stick’s blade position and response was proven to affect the skills of the hockey players \cite{lomond}. Studies \cite{brendan} even found the ideal zone for the blade to impact the ice for maximum shot speed and prove that puck speeds increase when impact happens closer to the toe of the blade. However, the effect of blade surface on performance has not been addressed so far, except for Frayne~et~al.~\cite{frayne} who concluded that the tape covering the blade used in the study did not have a positive effect on performance (when compared to bare blade).

To fill the gap in existing research, in this study we decided to focus on the influence of the physical parameters of the blade, such as hardness and associated damping properties and roughness on parameters of the shots, namely speed, rotation and energy. Several solutions for changing the blade surface exist and can be divided into two basic groups. The first one, further denoted as traditional tape, consists of a long strip of textile or similar material with glue on one side. This tape is then applied on the blade in a rather personalised manner by each player, differing in the parts of the blade covered, as well as the number of layers used. Within our study, from the following products: Howie’s Hockey Tape \cite{howies}, Renfrew Pro \cite{renfrew} and Comp-o-Stik \cite{compostick} we decided to select Renfrew Pro as a very popular product among the tested players.

Contrary to the traditional tapes, alternative blade covers are harder, rougher and are applied by sticking a single piece on the blade. From the spectrum of these products as BladeTape \cite{bladetape}, BladeShark \cite{bladeshark} and Specter Hockey Tape \cite{specter} we went for Specter as the one readily available. 

We have examined two types of shots that are most widely used in ice hockey, namely the Wrist shot and the Slap shot \cite{montgomery}. Wrist shot is quicker in release and players use it to surprise the goalkeeper, when shooting covered by a defender. The Slap shot is the most powerful shot with the highest speed, although it takes more time to carry out and allows the goalie to prepare. We have invited a set of $13$ players ($9$ lefties and $4$ righties) ranging from skilled amateurs through junior players to top NHL professionals. 
All players have carried out a series of Wrist shots and Slap shots. 

We have seen that for most players both speed and frequency of rotation of the puck where higher for both types of shots using Specter tape in comparison to Renfrew tape. Whereas for individual players the results varied depending on the type of the shot and other parameters, the collective statistics showed rather smooth results confirming the increase in both speed and frequency with almost certainty, while the expected increase was in the order of a few percent. We argue that this is due to the lower loss in damping, as well as due to the rougher surface of the Specter tape comparing to Renfrew. 


This paper is organised as follows: 
In Results we first present the data obtained from the experiments for individual players and then its collective statistical analysis. In the last part we elaborate on physical background of the results obtained. In Methods Section, we describe the setup, conditions and methodology of the experiments performed with both Renfrew and Specter as well as precision and pre-processing of data obtained for individual players. We also describe the experiment for measuring the friction coefficient. 


\section{Results}\label{chap:Results}
We analysed results for $13$ different players with various experience and skill level. 3 NHL professionals (Zdeno Ch\' ara, Tom\'a\v s Tatar, Martin Bako\v s) were joined by junior, senior and amateur hockey players with above average shooting skills. 12 players completed the experiment, 1 player was excluded because his stick broke during the experiment. 

\subsection{Individual players' statistics}
In the first step, statistical analysis was performed for each individual player. Expectantly, the stability of shots for each player was not perfect, due to unavoidable human influence. This is why we first decided to remove outliers from the raw data by arguing that some of the shots might have been spoiled.

Outliers were defined as lying outside Tukey's fences \cite{Wasserman2010}, i.e. outside of the interval
\begin{equation}  \label{eq_fences}
\centering
   [Q_1- 1.5 IQR, Q_3+ 1.5 IQR],
\end{equation}
where $Q_1$ and $Q_3$ are the first and third quartiles and IQR refers to the interquartile range:
\begin{equation}
    IQR=Q_3-Q_1.
\end{equation}

First we tested the level of correlation between the results in speed of the shots and frequency of rotation of the shots. The correlation coefficient for all available data (all players all shots) was at the level of $0.66$. This means that there is a reasonable, although not extremely strong, correlation between the two types of data. Based on this we decided to take each shot as a single event with two parameters (speed and frequency of rotation) and, if needed, to remove the shot as a whole, i.e. both the speed and the rotation from further analysis.


For each player independently both speed and frequency of rotation were evaluated for both types of shots and taping methods. 
To answer the question "How did the speed or frequency change using the Specter tape compared to using Renfrew tape?", relative gain (in per cent) was defined: 
\begin{align} \label{eq_gain}
    gain(X) &= \frac{X_{sticker}-X_{tape}}{X_{tape}}
\end{align}
where X is the parameter of interest, in our case average speed, average frequency of rotation or energy for each player and type of shot. Relative gain is a measure of impact of the taping method on shooting performance of each player: positive gain implies increase in performance with Specter compared to Renfrew, negative gain a decrease in performance.  

The effect of taping method on each player was thus characterised by two numbers: gain in speed and gain in rotation. To obtain a unifying parameter, we also calculated the gain in energy. Kinetic energy of a moving and rotating puck is defined as:
\begin{equation} \label{eq_energy}
    E = \frac{1}{2} m v^2 + \frac{1}{2} I \omega ^2,
\end{equation}
where $m$ is puck's mass, $I= \frac{1}{2} m R^2$ is puck's moment of inertia for rotation around its canonical axis, $R$ is puck's radius, $v$ is its speed, $\omega = 2 \pi f$ stands for angular speed and $f$ is rotation frequency.

Energy is interesting due to two reasons - firstly, it combines in a neat way both speed and frequency of rotation and characterises the shot by a single parameter. Secondly, the energy is, in the end, the physical entity that needs to be transferred from the player via the stick to the puck - shooting twice as fast needs four times more effort, as four times more energy is needed for such a shot.

\begin{figure}[t!]
        \centering
        \begin{subfigure}[b]{0.64\textwidth}
                \centering
                \caption{Speed of Wrist shots.}
                \includegraphics[width=\textwidth]{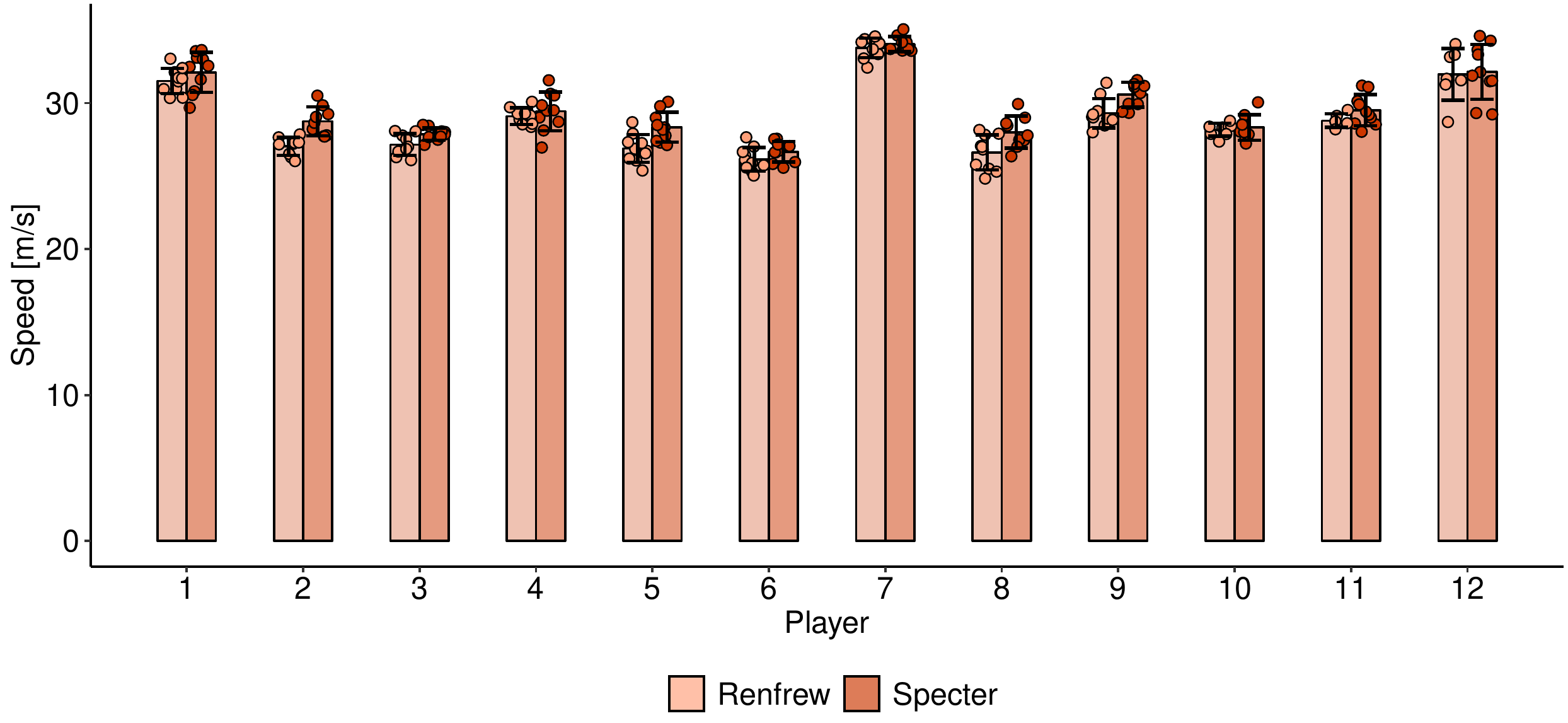}
                \label{Velocity_Wrist}
        \end{subfigure}
        \hfill
        \begin{subfigure}[b]{0.64\textwidth}
                \centering
                \caption{Speed of Slap shots.}
                \includegraphics[width=\textwidth]{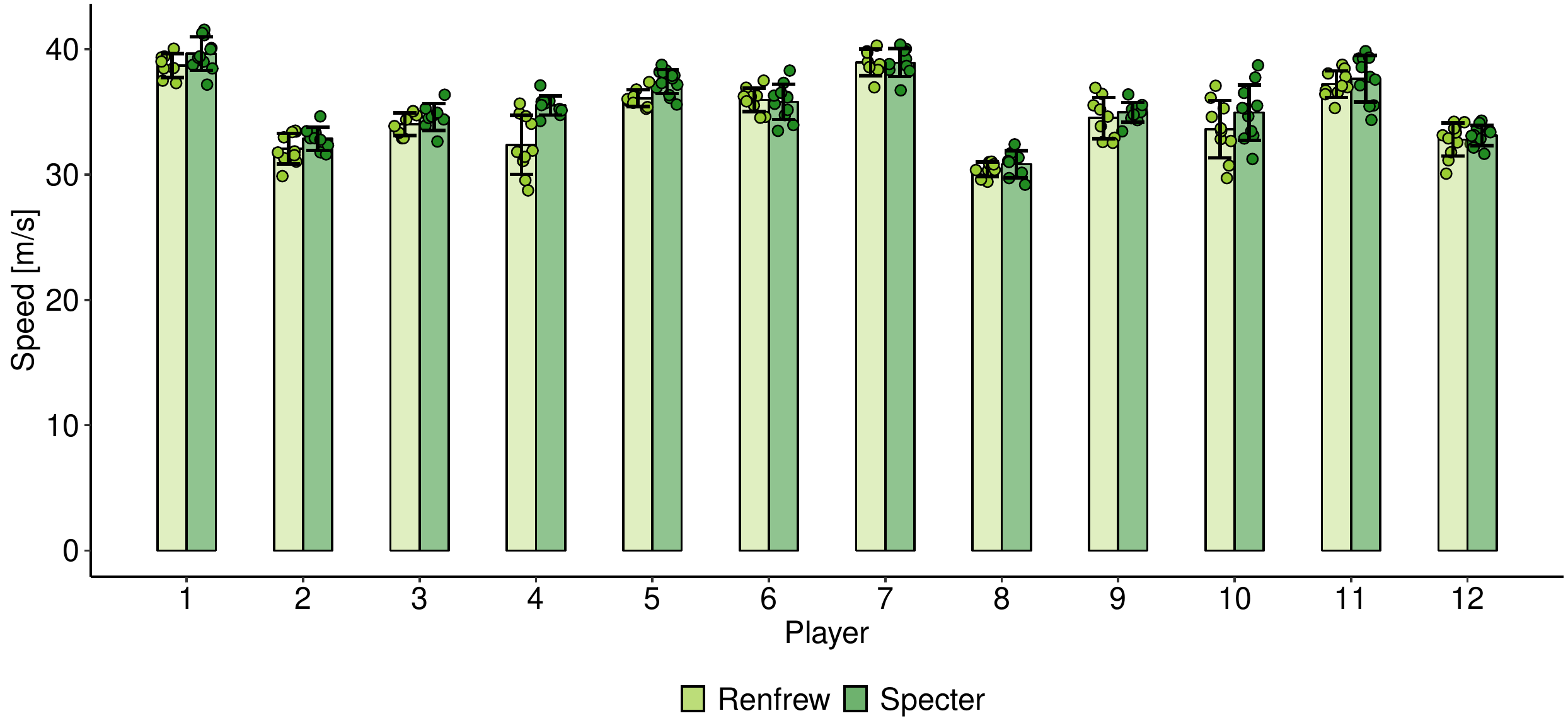}
                \label{Velocity_Slap}
        \end{subfigure}%
        \caption{\emph{Speed.} Figure shows speed of \textbf{(a)} Wrist shots and \textbf{(b)} Slap shots for individual players. Bars represent mean value, error bars standard error and dots correspond to individual shots. For Wrist shots, average values are consistently higher for Specter tape. For Slap shots, the average values are clearly higher for a couple of players, in one case the value is slightly smaller and in one case almost identical.}
        \label{fig:speed}
\end{figure}

\begin{figure}[h]
        \centering
        \begin{subfigure}[b]{0.64\textwidth}
                \centering
                \caption{Rotation frequency of Wrist shots.}
                \includegraphics[width=\textwidth]{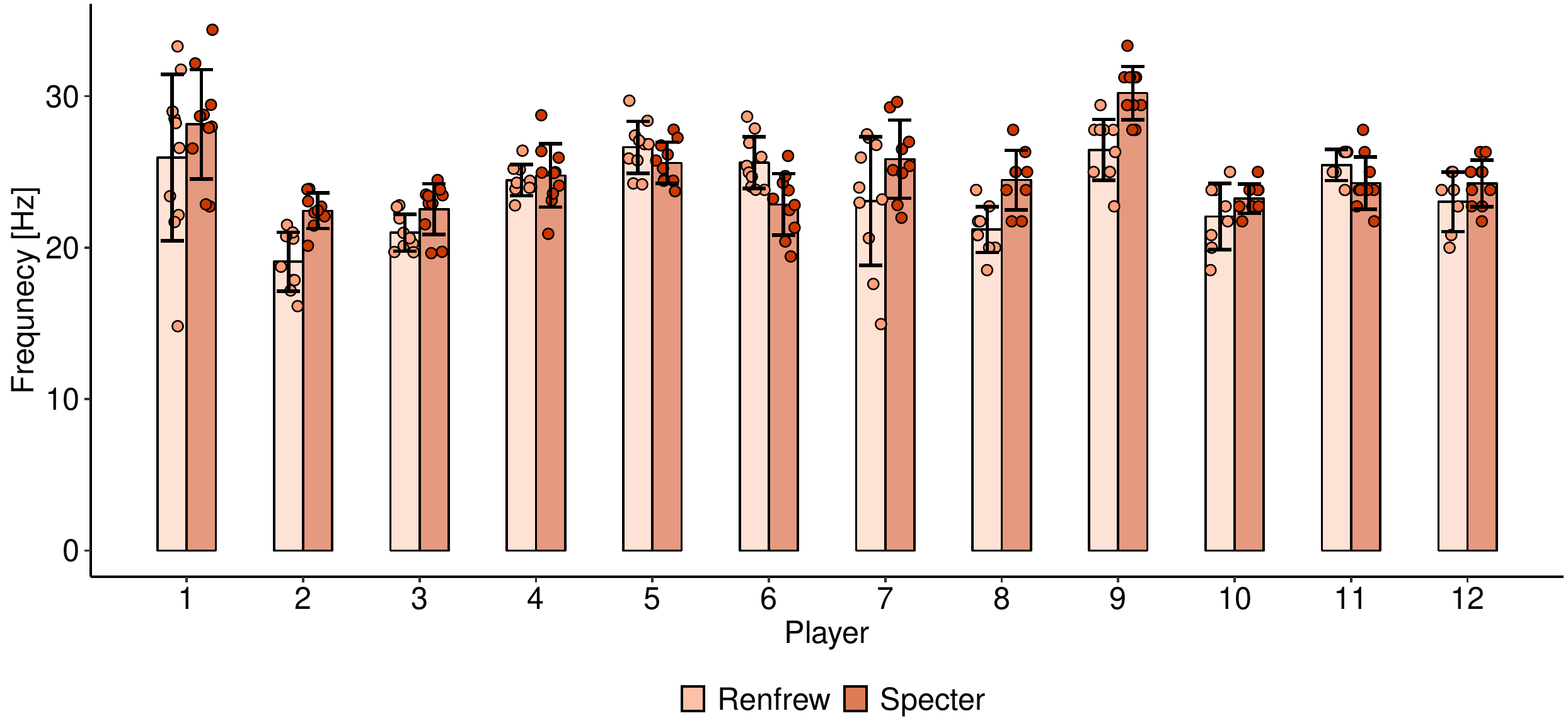}
                \label{Rotation_Wrist}
        \end{subfigure}
        \hfill
        \begin{subfigure}[b]{0.64\textwidth}
                \centering
                \caption{Rotation frequency of Slap shots.}
                \includegraphics[width=\textwidth]{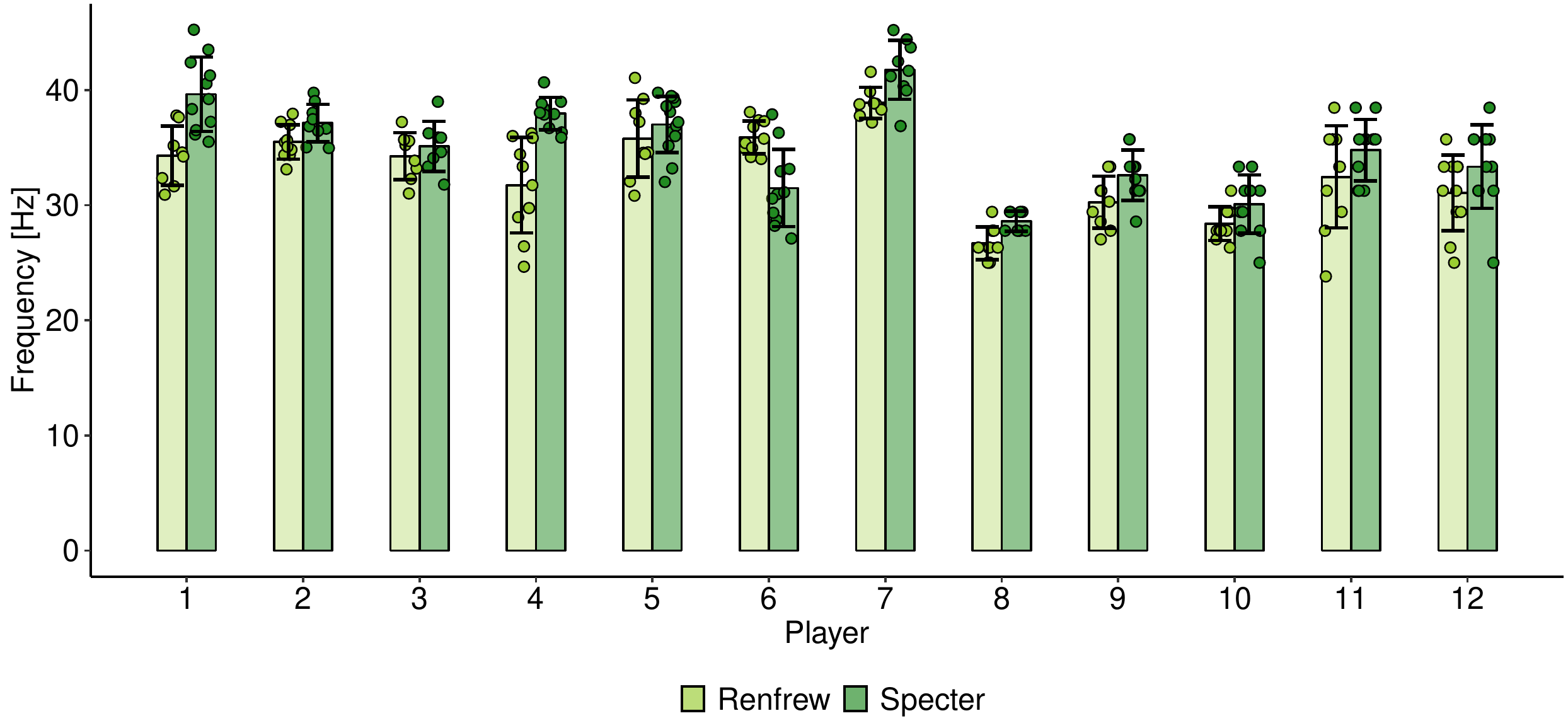}
                \label{Rotation_Slap}
        \end{subfigure}%
        \caption{\emph{Rotation frequency.} Figure shows rotation frequency of \textbf{(a)} Wrist shots and \textbf{(b)} Slap shots for individual players for Renfrew tape and Specter. Bars represent mean value, error bars standard error and dots correspond to individual shots. For Wrist shots, the average values are rising consistently more than for speeds, but there are two players with a clear drop. The smaller stability of the shots led to higher standard deviations, thus the rise in this case is also comparable or smaller than the standard deviation except of one player. For Slap shots, the average values are higher for all but one player.}
        \label{fig:rotation}
\end{figure}

\begin{figure}[p]
        \centering
        \begin{subfigure}[b]{0.7\textwidth}
                \centering
                \caption{Gain in speed}
                \includegraphics[width=\textwidth]{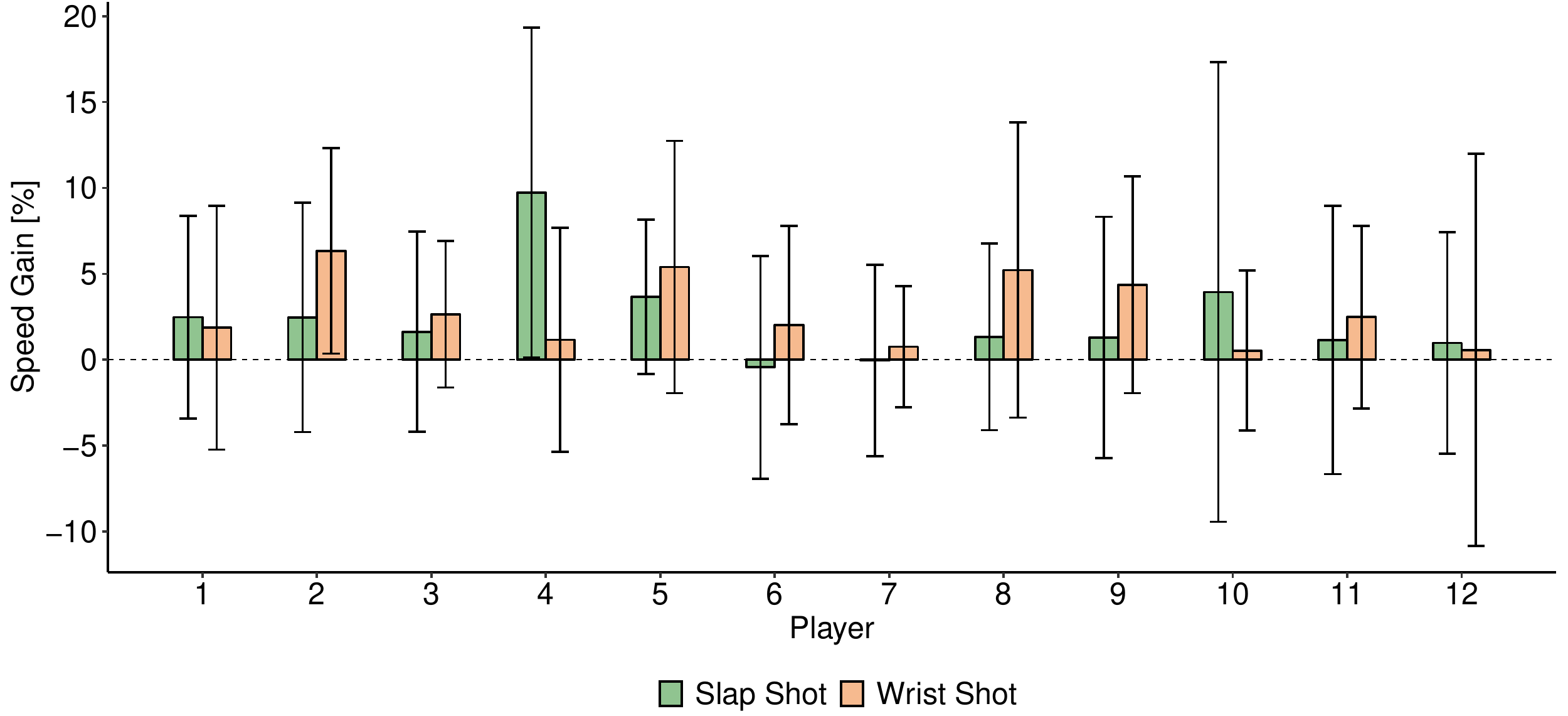}
                \label{Velocity_gain}
        \end{subfigure}
        \vfill
        \begin{subfigure}[b]{0.7\textwidth}
                \centering
                \caption{Gain in rotation frequency}
                \includegraphics[width=\textwidth]{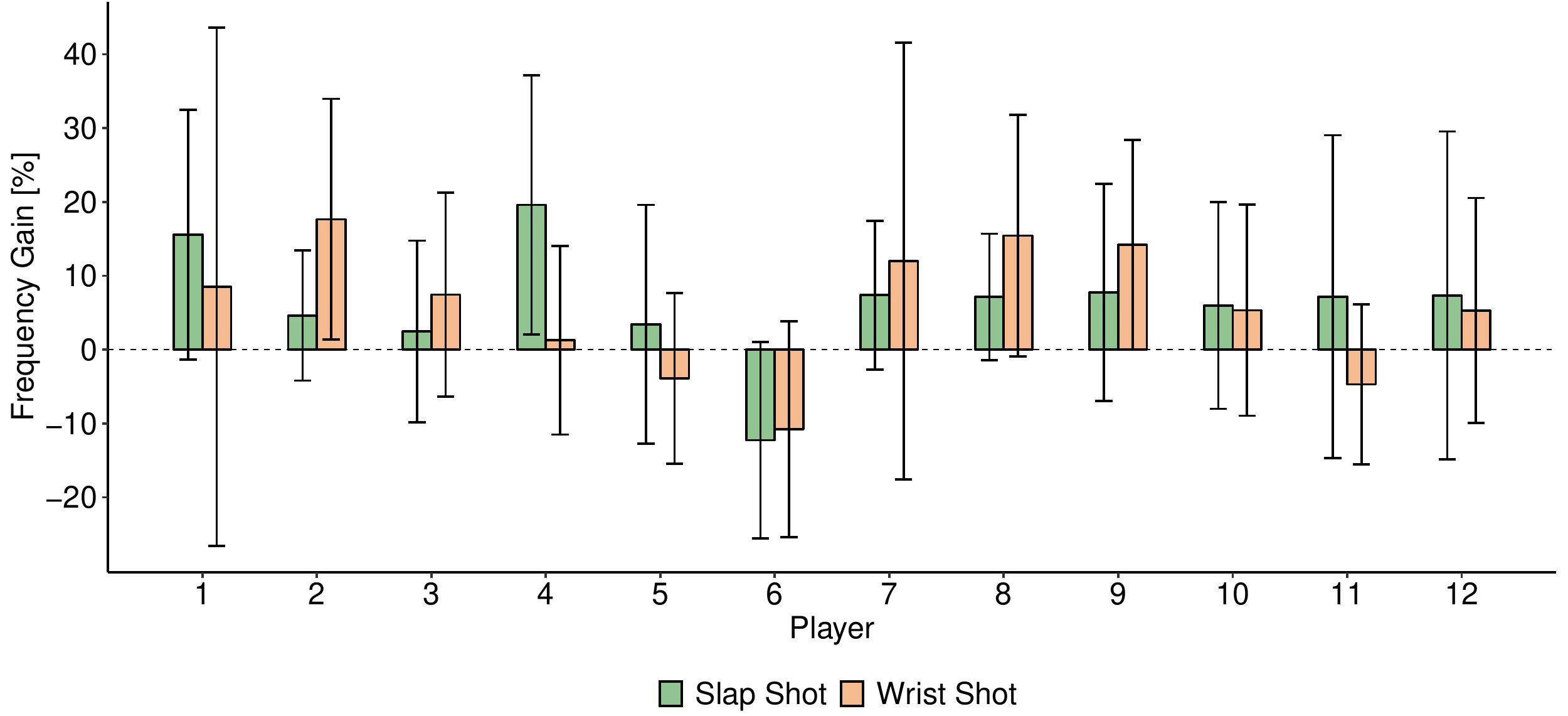}
                \label{Rotation_gain}
        \end{subfigure}%
        
        \begin{subfigure}[b]{0.7\textwidth}
                \centering
                \caption{Gain in energy}
                \includegraphics[width=\textwidth]{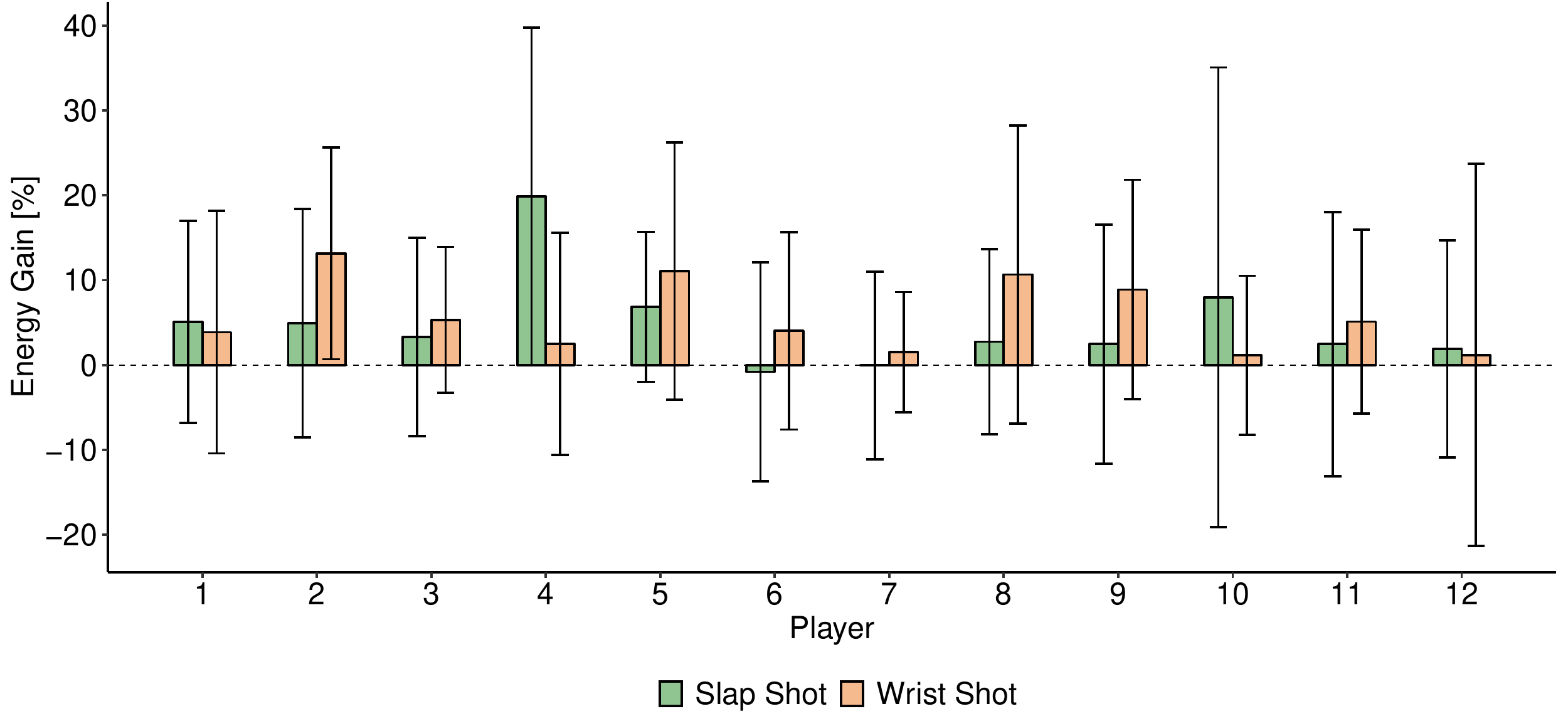}
                \label{Energy_gain}
        \end{subfigure}%
        \vfill
        \caption{Individual gains in speed, rotation and energy for Slap and Wrist shots. Except for three players, the gains in all parameters were always positive. Some players reached gains higher than $10\%$, which is a considerable increase in performance.}
\end{figure}

\subsection{Speed, Rotation and Energy}

We analysed how speed and rotation of the shot differ between the two taping methods. The results for speed are shown in Figure \ref{fig:speed} and the results for rotation in Figure ~\ref{fig:rotation}. The gains in speed and rotation are show in Figures \ref{Velocity_gain} and \ref{Rotation_gain}, respectively. 
The results obtained show consistent increase in both speed and rotation, albeit comparable or smaller than the standard deviation for individual players. Both the mean and the variance in rotation gain are larger than of gain in speed, presumably due to lower stability of the shots for most of the players.

Speed and rotation can be neatly joined into a single parameter - the energy of the shot, as defined in (\ref{eq_energy}). The gains in energy are shown in Figure~\ref{Energy_gain}. In almost all cases the gain is positive and in some cases even relatively high: four players achieved the gain of more than $10\%$ and one of them even almost $20\%$. The energy gains -- both averages and standard deviation -- are in all cases very close the twice the values for speeds. This is due to the fact that, as expected, the energy is mostly determined by the speed of the puck and the speed enters the energy in second order.

\begin{figure}[t]
        \centering
        \begin{subfigure}[b]{0.33\textwidth}
                \centering
                \includegraphics[width=\textwidth]{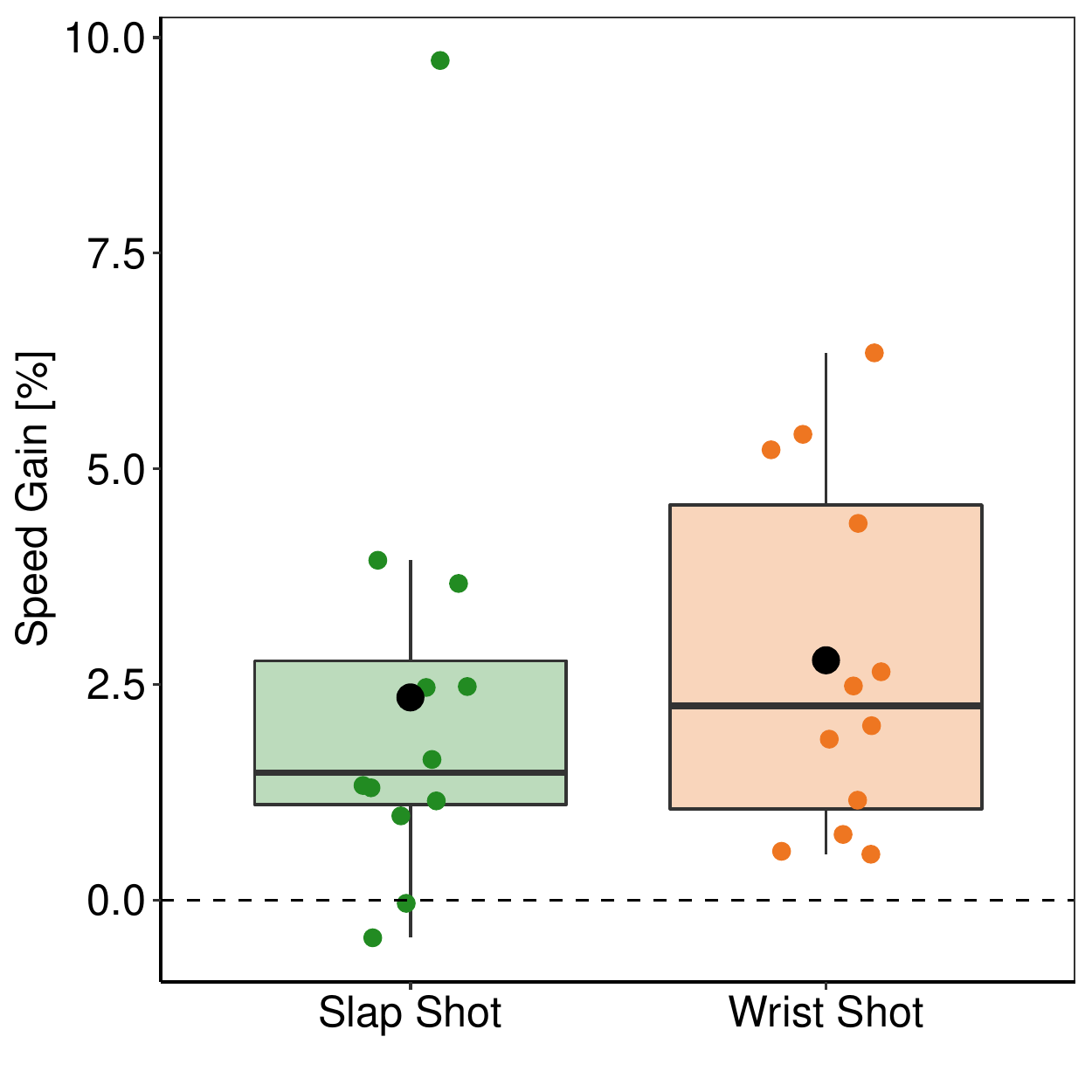}
                \caption{Gain in speed}
                \label{fig.velocity_gain}
        \end{subfigure}
        \hfill
        \begin{subfigure}[b]{0.33\textwidth}
                \centering
                \includegraphics[width=\textwidth]{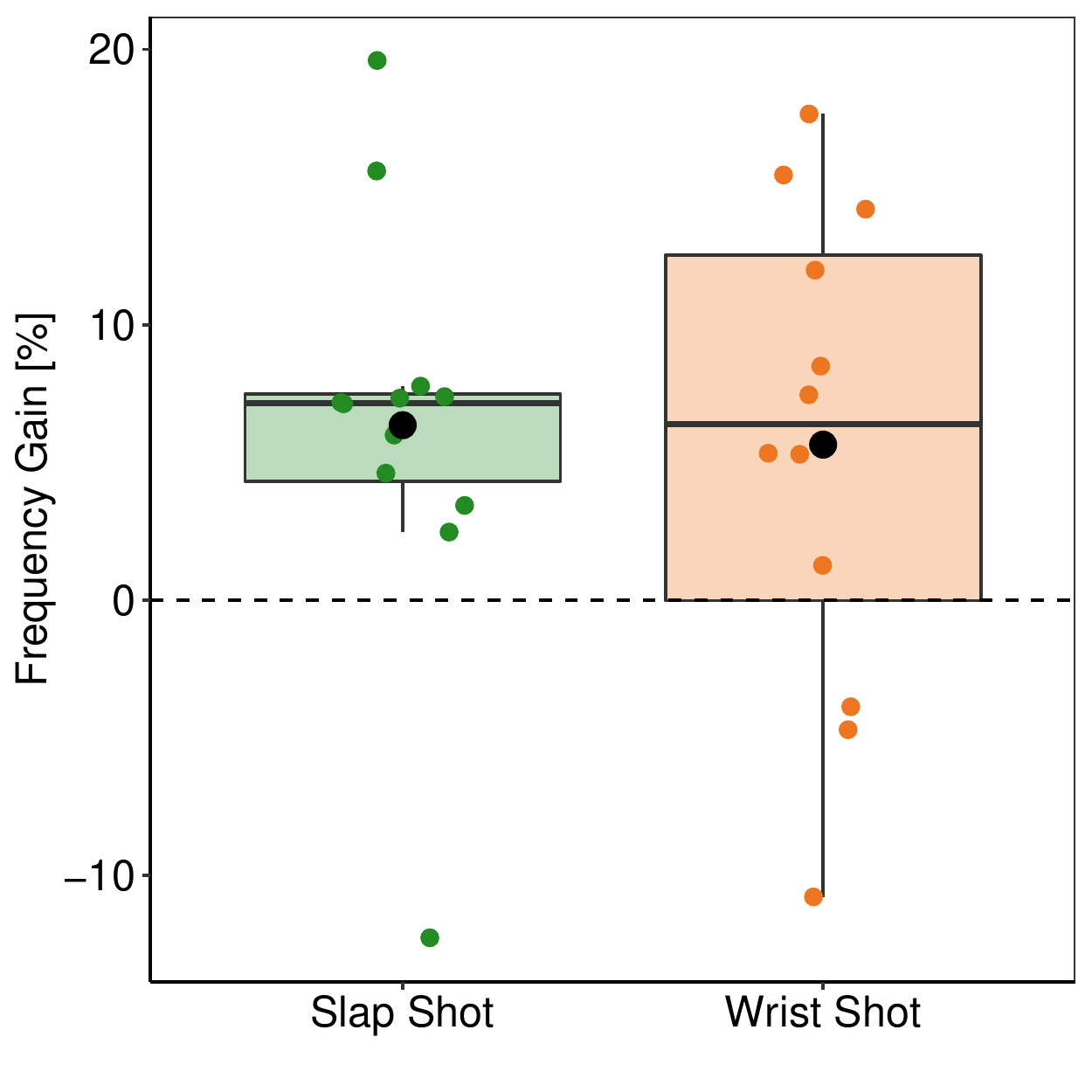}
                \caption{Gain in rotation frequency}
                \label{fig.rotation_gain}
        \end{subfigure}%
        \hfill
        \begin{subfigure}[b]{0.33\textwidth}
                \centering
                \includegraphics[width=\textwidth]{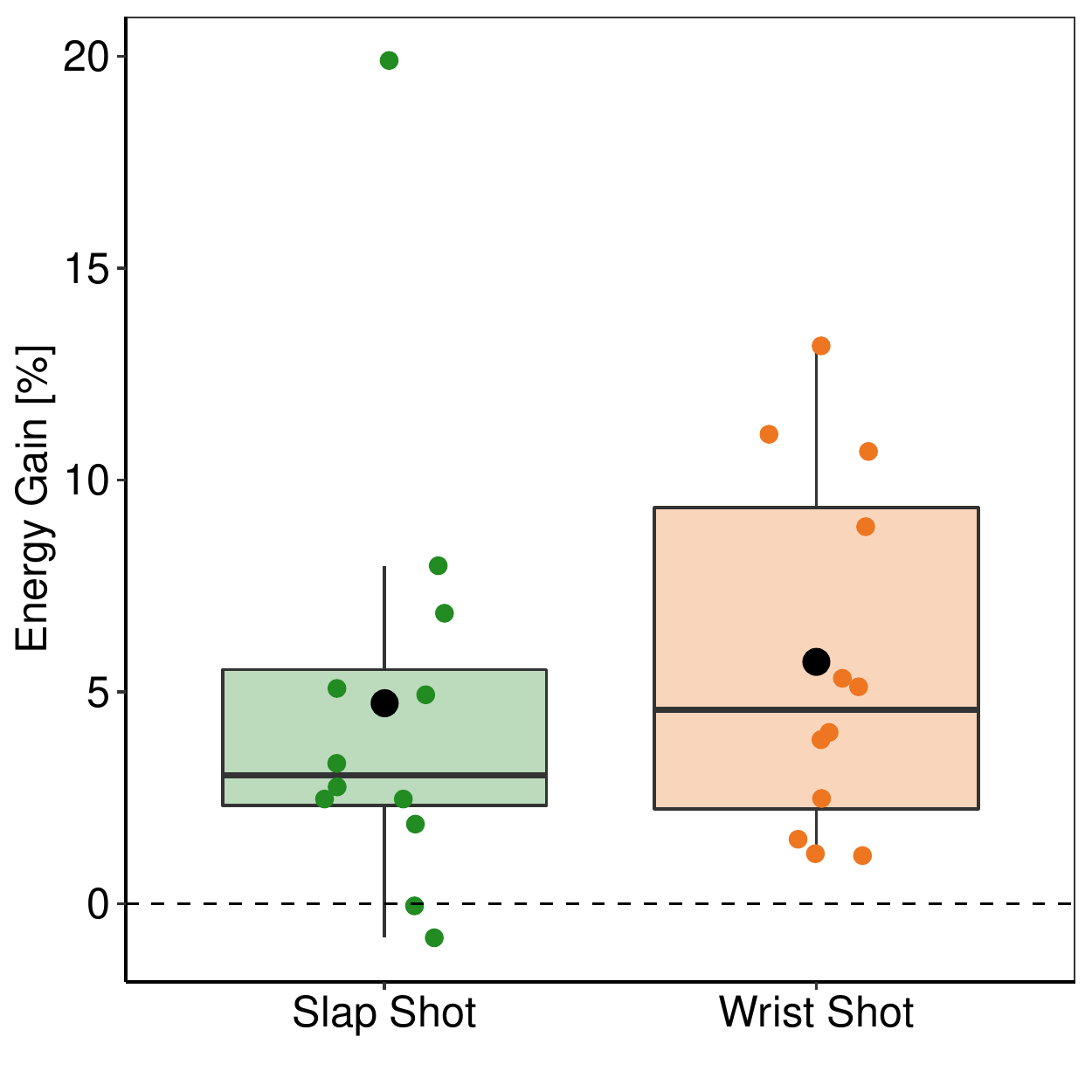}
                \caption{Gain in energy}
                \label{fig.energy_gain}
        \end{subfigure}%
        
        \caption{Boxplots of gains in the three performance parameters for both types of shots: Slap Shot (green) and Wrist Shot (orange). Median (horizontal thick line) and mean (full circle) as well as individual data points are shown.}
\end{figure}

\begin{table}[h]
\centering
\begin{tabular}{|l|c|c|c|c|}
\hline
         & Average gain  & \multicolumn{1}{c|}{\begin{tabular}[c]{@{}c@{}}Minimal gain \\ 95\% confidence\end{tabular}} & \multicolumn{1}{c|}{\begin{tabular}[c]{@{}c@{}}Minimal gain \\ 90\% confidence\end{tabular}} &\multicolumn{1}{c|}{\begin{tabular}[c]{@{}c@{}}Probability of \\positive gain \end{tabular}} \\ \hline
Speed &        $2.78 \%$               &       $1.71\%$       &       $1.97\%$       &   $99.97\% $                                                                                     \\ \hline
Rotation &         $5.64\%$              &         $1.08\%$      &       $2.18\%$       &   $97.59\%$                                                                                   \\ \hline
Energy   &         $5.71\%$              &          $3.52\%$    &       $4.05\%$        &   $99.97\% $                                                                          \\ \hline
\end{tabular}
\caption{ \emph{Gains of Specter tape for Wrist shots.} Table shows the gain of Specter tape for Wrist shots. One can conclude that with confidence of more than $99,95\%$ the Specter tape leads to a gain in speed and energy. The confidence in rotation is slightly smaller, albeit the expected gain is higher, due to larger variance in results.}
\label{Student_Wrist}
\end{table}
\begin{table}[!ht]
\centering
\begin{tabular}{|l| c| c|}
\hline
         & \multicolumn{1}{c|}{Average gain} & \multicolumn{1}{c|}{\begin{tabular}[c]{@{}c@{}}Probability of positive gain \end{tabular}}\\ \hline
Speed &        $2.35\%$                 &  $99.88\% $    \\ \hline
Rotation &         $6.35\%$               & $98.95\% $    \\ \hline
Energy   &         $4.73\%$            &   $99.73\% $      \\ \hline
\end{tabular}
\caption{ \emph{Gains of Specter tape for Slap shots.}
Table shows the gain of Specter tape for Slap shots. One can conclude that with confidence almost $99\%$ that Specter tape leads to a gain in all parameters. One can also see that while the gain in speed and energy is slightly smaller compared to Wrist shots, the gain in rotation is higher.}
 
\label{Student_Slap}
\end{table}

\subsection{Collective statistics}
Expectantly, the results for individual players are not particularly stable. Firstly, because each player has his own technique, that might be more or less adapted for using a different taping method. Secondly, having about $10$ shots per player is statistically not enough to draw definite conclusions for individual players. Nevertheless, we decided to keep the total number of shots per session at 40 because of possible tiredness of players with higher load. In this setting, we can be certain that with regular breaks between shots tiredness will not be a factor. 


Results for collective statistics can be visualised in a form of boxplots (Figures~\ref{fig.velocity_gain},~\ref{fig.rotation_gain} and~\ref{fig.energy_gain}). In each boxplot, median (horizontal line) and mean (full circle) of the sample are shown. The box encloses the middle $50\%$ of data, i.e. data lying between the first and third quartile (lower and upper hinge). Vertical lines point to inner fences, as defined by (\ref{eq_fences}). Coloured small circles are individual data points, corresponding to each boxplot (value of gain for each player). Dashed horizontal line denotes zero gain.  
The plots suggest that the gain is consistently higher than zero for all parameters. In other words, players performed better when using Specter compared to using Renfrew tape.

To determine whether the taping method has significant influence on the parameters of shots, we performed statistical tests on gains. Depending on whether the sample analysed is likely from a normal distribution, two types of tests are used: Student's t-test for normally distributed data and Wilcoxon signed rank test for samples, which do not fulfil the condition of normality. In our case, Wrist shots were analysed using Student's t-test whereas for Slap shots, Wilcoxon signed rank test was used. Unlike for Wilcoxon test, output of t-test also includes confidence intervals, therefore the $90\%$ and $95\%$ confidence intervals have been calculated for Wrist shots. The results are shown in Tables~\ref{Student_Wrist} and \ref{Student_Slap}. Details of statistical analysis are provided in Methods.

Several conclusions can be drawn here. The first and most obvious one is that with very high confidence (higher than $97\%$ for all cases) the Specter tape enhances player's performance in all relevant parameters - that means a player will achieve higher speed, rotation frequency and energy of the shot using Specter tape when compared to Renfrew tape.
Furthermore, while the average gain in rotation is higher than in speed (in fact more than double), the associated uncertainty is higher as well. This leads to the fact that the gain expected with very high probability is smaller than that for speed.


Interestingly, the variation between players in rotation frequency of Wrist shots was much larger than for Slap shots, though the average and median gains are similar. Also, judging from individual statistics (Fig.~\ref{fig.rotation_gain}), Wrist shot rotation gain for each player is less stable than Slap shot rotation gain. This may be due to the different technique of the shots, that will be analysed in detail in the next subsection.


\subsection{Physical background}

As described in previous subsections, it is clear from the experiments that specifically developed blade stickers like Specter lead to better results in both speed and rotation than standard tapes, such as Renfrew. Here we analyse 
the underlying physical principles. 


The naive interpretation of a Slap shot is that it can be compared to a slash of a hammer on a nail, where in a collision the momentum and energy is transferred from the stick to the puck. This is, however, true only to a certain extent. The blade speed of the Slap shot goes through several phases  \cite{kays}. Making use of rather high elasticity of the stick, it is leaned against the ice (or a shooting mat as in our experimental setup) and bent like an arch. During this phase energy is accumulated in the elastic energy of the stick. The blade then touches the puck and collides in a mostly inelastic manner, while accelerating the puck to about $60\%$ of its final speed \cite{kays}. Then the energy accumulated in the bending of the stick is gradually released as the blade is freed from the surface. 
Whereas the time of acceleration of the stick is on the order of tens of milliseconds (confirmed both by \cite{kays} and our measurements), the time of the primal collision of the blade and puck is much shorter, certainly less then a millisecond (below time resolution in both studies). 

For Wrist shots, the situation is different. Here the blade touches the puck from the very beginning, so there is no collision involved. The accumulation of energy in bending is used to a smaller extent and the acceleration of the puck takes much longer (about $100ms$ or more). The energy and momentum transfer from the player to the puck is thus more direct meaning the forces acting on the puck are mostly limited to forces the player themselves can expend, which requires longer contact. During this time, the puck rotates along the blade, allowing both better aiming and stability. 

Let us first estimate the forces and momenta transferred during shooting. For the case of speed of $30ms^{-1}$ and the mass of the puck $165g$, the momentum transferred is in the order of $5kgms^{-1}$. For Wrist shots, we can rather safely expect the force to be constant during contact (approximately $100ms$), resulting into a force of $50N$ 
. 
For Slap shots the force is highest during the collision and much smaller during the further acceleration later on. If we approximate the momentum transferred during the collision lasting  $0,5ms$ to be $3kgms^{-1}$, we end up with a force of $6kN$ or more, as the collision might take even much shorter time. 

\subsubsection{Damping}
While Renfrew is basically a textile tape with glue on one side, Specter has a more elaborate composition of several different layers. The thickness of both of them is very similar: \(0,37mm\) for Specter \cite{S_width} and \(0,254mm\) for Renfrew \cite{R_width}, but the elastic/damping properties are substantially different. While Specter does not deform almost at all even under very high pressures, Renfrew can be easily deformed even by hand (e.g. by pressing the disk of unused tape). 

We believe that the loss of energy for Renfrew (from now on, we shall switch the viewpoint and speak about loss of Renfrew rather then of gain of Specter) is caused by the inelastic deformation of the tape. To elaborate on this, let us calculate the energy balance.  

The energy of the puck is in the order of $100$ Joules and the experimentally confirmed loss of Renfrew is thus in the order of $5$ Joules (see Tables \ref{Student_Wrist}, \ref{Student_Slap}). For Slap shots, if we take  $6kN$ as the force, it corresponds to the deformation in thickness of about $0.8mm$. This is not realistic, as it is slightly more than the thickness of two layers of the tape as used during our experiment. On the other hand, as all the numbers above are gross estimates where a factor of two can certainly enter (especially via the duration of the collision, which could surely be shorter), we can conclude that the hypothesis does not contradict experimental data.   

The situation is rather different for Wrist shots. Here the puck does not touch the blade at one contact point, but rather rolls on it during the shooting. Thus while the forces acting here are much lower (as there is no direct collision involved), the deformation of the tape has a higher effect with losses distributed on a large portion of the blade in the form of rolling friction. 
So while the primary source of losses (damping in the tape) is the same, the mechanism is substantially different and harder to estimate quantitatively. 

Let us model the experiment by rolling resistance of the puck rolling on the blade. 
The force resisting rolling can be modelled as 
\begin{equation}
   F_R=\sqrt{\frac{z}{d}}F_N, 
\end{equation}
where $z$ is the deformation depth, $d$ is the diameter of the puck ($76.2 mm$) and $F_N$ the normal force. We can express the energy loss as 
\begin{equation}
   E_L=F_R s = \sqrt{\frac{z}{d}}F_N s, 
\end{equation}
where $s$ is the path travelled by the puck. 
Taking $F_N=50N$, the path along the blade $20cm$ and the deformation depth $0.4mm$, we end up with an energy loss of around $0.7J$. This is certainly less than the experimentally seen loss in order of $3.6J$ (see Tables \ref{Student_Wrist}, \ref{Student_Slap}). We conclude that the rolling friction connected with deformation of the Renfrew tape has a non-negligible effect on the losses, but most likely it is not the prevailing source. We discuss other possible source qualitatively below.  

\subsubsection{Friction}
Naively one would expect the gain in rotation to be approximately the same as the gain in speed, based on a simple scaling principle - if mechanism of the shot with Specter are similar but quicker than with Renfrew, the speed should differ to the same extent as rotation. 

This was, however, not the case in our experiments - the average gain in rotation was almost double for Wrist shots and almost triple for Slap shots compared to gain in speed. At the same time, the gain in rotation was higher than in speed for 9 out of 12 players for Wrist shots and 10 out of 12 for Slap shots. To understand this, we again look at the shooting style of the two types of shots. 

Here the understanding seems to be simpler for Wrist shots: the puck is rolling on the blade during the shot and therefore also gains rotation. The longitudinal force responsible for rotating the puck is, however, strictly limited by the friction force, which is proportional to the normal force and the friction coefficient. There is a large difference in friction coefficient between the two tapes: while Renfrew is basically textile, Specter has a surface similar to sandpaper with a clearly rougher surface. We have measured the friction coefficients for both Specter ($f=0.69\pm 0.02$) and Renfrew for different configurations (longitudinally $f=0.46\pm0.07$, transversely "downstairs" $0.43\pm0.08$ and transversely "upstairs"  $0.50\pm0.11$. The configuration used for measuring the shots was upstairs; for details see the Methods section). While the different configurations of Renfrew tape do not differ significantly in friction coefficient, they are all much smaller than the one of Specter. Thus the magnitudes of longitudinal forces caused by friction while the puck is sliding on the blade are increased even more when using Specter than when using Renfrew compared to the effective normal (accelerating) forces, as both the normal force and the friction coefficient are increased.  

For Slap shot, the shot can be modelled as a single hit into the puck followed by a gradual acceleration. While the second phase of the movement can be described in a similar manner as for Wrist shots, the first one is substantially different. Here the degree of rotation is given mostly by the level of non-centrality of the hit - the further off-axis the hit is made (the more the velocity vector of the blade and normal vector of the surface of the blade differ), the more rotation for a given speed of the puck is achieved. This level is basically determined by the level of tilt of the blade as well as the exact point of contact, as the blade is curved. Neither of these is easy to control, what could partly explain why the variation in rotation for different players was much higher compared to variation is speed. 

Like for Wrist shots, the gain in rotation for Slap shots is on average higher than the gain in speed. Here one can argue that for Specter, with higher friction coefficient, the longitudinal force acting on the puck is generally higher than for Renfrew for the same normal force. For close-to-central hits, which is the case for Slap shots (as players aim to shoot with the largest speed possible), longitudinal forces contribute mainly to the gain in rotation. Thus higher longitudinal forces (due to higher friction) further contribute to the increase in rotation. For a more detailed quantitative analysis one would need to look in detail on the shot itself. This was not possible with the given equipment and goes beyond the scope of this research. 


We also believe that the increased friction coefficient of Specter might contribute to a higher shooting speed in comparison to Renfrew for Wrist shots. During the shot, at certain moment the blade moves above the puck and looses the contact and hence the ability to further accelerate the puck. With higher friction, longer contact between the blade and the puck can be maintained, as the contact point will be more efficiently elevated by the blade while the centre of the puck moves further mostly horizontally. A mere $1cm$ longer contact with the blade would lead (in the constant force approximation) to more than $1\%$ increase in the total energy. For a quantitative analysis of this hypothesis one would need to measure the contact time between the puck and the blade with very high precision and this was far above our experimental possibilities. 

To summarise, we have shown that the loss of Renfrew in speed is partly due to higher energy losses via damping in the tape. Damping itself can mostly explain the loss in speed for Slap shots, as well as around $20\%$ of the loss in Wrist shots. The rest of the difference can be attributed to the significantly higher friction coefficient of Specter, which allows for longer contact between the blade and the puck. This results in higher release speed of the puck, as the puck is being accelerated longer. The gain in rotation is even larger than gain is speed for both types of shot; this can be explained by the fact that longitudinal forces are increased both due to larger transversal forces (due to smaller damping), as well as due to higher friction coefficient. 


\section{Discussion and Conclusions}\label{chap:Discussion}

We have analysed the influence of the physical properties of the ice-hockey stick blade surface on both speed and frequency of the rotation of the puck. The analysis was performed by taking high-speed video recordings of shots of $13$ different hockey players of different levels of skills and analysing it via a tracker software. We have also calculated the energy of the shots as a unifying parameter for both speed and frequency of rotation.

We have seen that due to instability of the individual shots, it is rather hard to draw any conclusions for individual players. Each of them shot at least $10$ shots both with the Renfrew tape and Specter tape, both by Wrist shot and Slap shot. It was not possible to increase the number of shots by one player due to different reasons - if doing more at once, it would lead to increased tiredness and decrease of performance. On the other hand doing it in more shifts would make it complicated to guarantee the same outer conditions, in particular the overall performance level of the player.

With these limited number of shots we have seen that for most cases, i.e. almost all players for all or almost all parameters, the harder and rougher surfaced achieved by applying of the Specter tape led to increase in performance: the shots were quicker and had higher frequency of rotation, which made them more stable on one side and harder to track by the goalie on the other. In most cases the change for individual players was smaller than the standard deviation. 

This is why we also performed collective statistical analysis. 
As parameters of shots vary greatly from player to player, we compared the changes as the unifying parameters and performed statistical tests on them (Student's t-test and Wilcoxon signed rank test). We have seen that with very high confidence one can conclude that the harder and rougher surface leads to higher speed, rotation and energy of the shots. The expected gain is roughly $2.5\%$ in speed, which translates (depending on the player) to about $0.75$ m/s or almost $3$ km/h. For Wrist shots, when shooting from $10$ meters, it makes a difference of about $28cm$ in the shot trajectory, which is comparable to the diameter of the goalie catching glove. For Slap shots, when shooting from the blue line ($17,3$ meters), the average difference is as large as $40$ cm. 

We argue that the results can be explained by two basic differences between surfaces achieved by applying Renfrew and Specter tape respectively. The first one is softer, what causes higher damping leading to smaller speeds and rotational frequencies. The later has also a rougher surface, which allows higher rotations even for the same speed, as well as longer contact between the blade and the puck. More detailed analysis of the shots itself would be needed for a more complex analysis, but this was beyond possibilities of our experimental equipment. 


\section{Methods}\label{chap:Methods}

\subsection{Experimental methods}

The primary motivation for setting up the experiment was to guarantee the most stable conditions for shooting players. Use of own equipment (stick, gloves), stable shooting conditions (shooting mat), stable room temperature and lighting guaranteed as stable experimental conditions as possible. Players were informed about the purpose of the study and voluntarily participated in it as expert collaborators; template of the Consent to Participate which was signed by the players is provided as supplementary material. A written confirmation of compliance on ethical standards based on the procedures as stipulated by the Ethical code issued by the Ethical Committee of the Slovak Academy of Sciences is also provided as supplement. 

\begin{figure}[h]
\centering
\includegraphics[width=.7\textwidth]{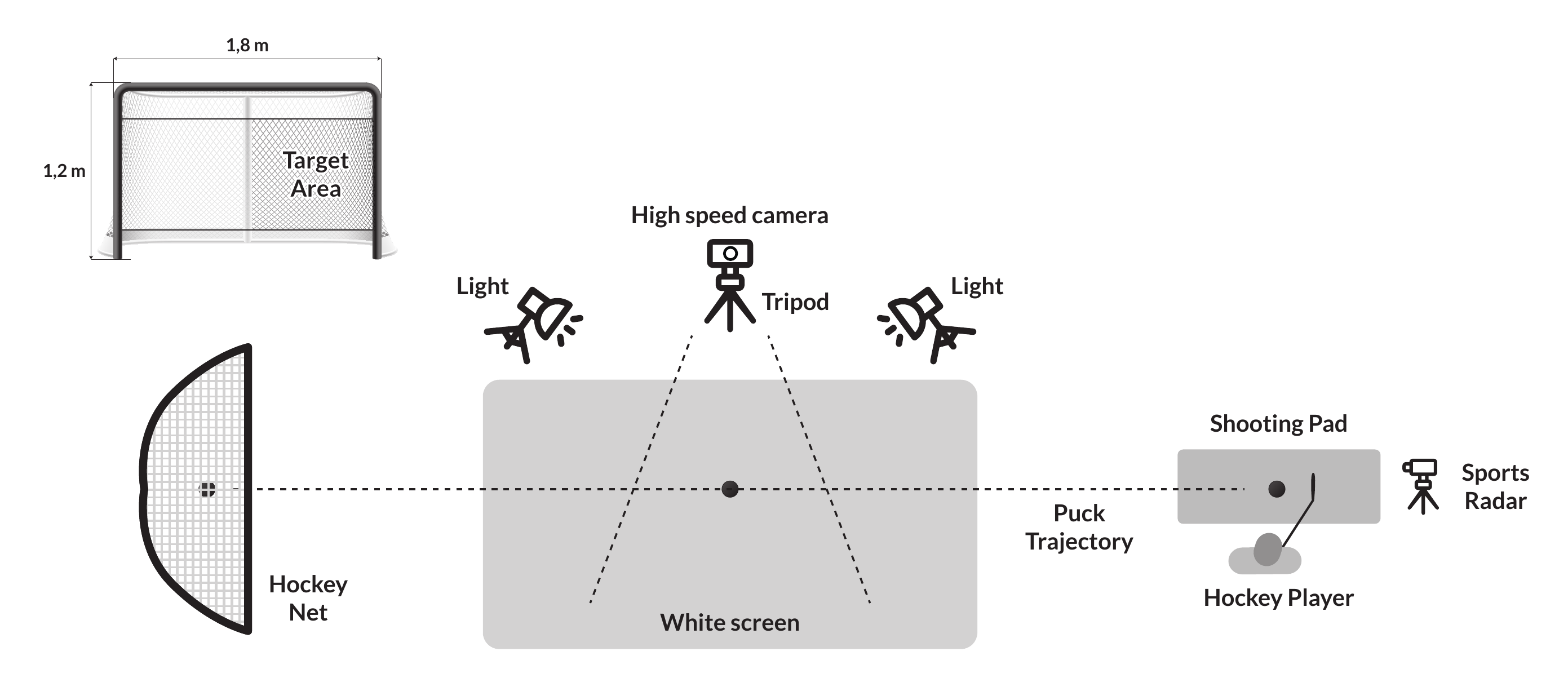}
\caption{\emph{Sketch of the experimental situation}}
\label{Sketch}
\end{figure}

Experiments were carried out in the Hockey Development Centre \cite{HDC2018}. 
Players were shooting from stand and the puck was placed on a shooting mat, simulating conditions on ice. Distance to goal was approximately $6$ meters, as described in Figure~\ref{Sketch}. Only those shots were taken into account that fitted into a small region of the goal marked by strings and spanning. The defined part of the net was a rectangle of $90 cm x 82 cm$ placed $20 cm$ above the surface. Taking into account the findings of Paquette et al.~\cite{yannick}, the defined part did not include the top of the net. We decided not to include the bottom part of the net either. Only shots, where the puck was not sliding on the surface were considered successful, to prevent possible influence of the surface. 
Players were expected to shoot four series of $10$ successful shots first using Renfrew tape for Wrist and Slap shots and then using Specter sticker, again for both Wrist and Slap shots. The overall efficiency of shots was more than $80\%$ and the maximum number of shots needed for a single series was $16$.
Renfrew tape wrapped from heel to toe in two partially overlapping layers on player's blade, then the tape was removed and on the same stick Specter tape was applied. After a break of $10$ minutes a series of the same shots was carried out. The hockey sticks used by the players were the following brands: CCM \cite{ccm}, Bauer and Easton \cite{bauer}, Sher-wood \cite{sherwood} and Warrior \cite{warrior}.
The whole experiment with one player took about $45$ minutes, including the break. During this time period each player shot roughly $50$ shots, what is comparable to a usual one hour training on ice, where further exercises are carried out as well. We haven't seen any statistically significant correlation between the time (order) of the shots and the speed or rotation frequency, so we can conclude that any effects connected with tiredness have only negligible influence on the results. 

Shots were recorded about $2$ meters from the release point with a digital high speed camera Sony RX100-V \cite{sony} on full HD resolution and $1000$ fps. Three to four coloured dots (depending on the run) of different colours were positioned on each puck, one in the middle and the rest on the edge. Pucks were only coloured with paint not to influence the weight of the puck and to provide a smooth shooting motion for the players. Videos were down-sampled by a factor of four and analysed by a tracker software. For some particular runs the software was not able to provide results due to low visibility of one of the dots or because of puck topple; in such cases the recording was analysed frame by frame by hand. 
Depending on the actual speed, about $30$ frames were captured for each shot and on each frame the size of the puck (measured as the distance between centres of points on the edge) was about $35$ pixels. This allowed to see about one period of rotation of the disc which was enough to calculate speed, rotation and energy of the individual shots.

\subsection{Precision}
An area of about $1$ meter long was covered by the camera with $1980$ pixels, leading to the resolution of about $0.5$ mm per pixel. In post-processing the resolution was decreased for the tracker software by a factor of 2 in each axis, changing the resolution to about $1$ mm per pixel.

The size of the dot on the puck was about $1$ cm in diameter with a precision of up to $1$ mm, making it comparable to the pixel size. This led to a precision in speed of almost $0.1\%$, much better than the differences between different shots of the same type of the same player.

Rotation was read out using a series of angles formed by two dots placed closed to the border of the puck. With $76.2$ mm diameter of the puck the distance between centres of the dots was about $50$ mm, leading to a precision of $2-4\%$ in the angle. This is substantially less than the speed itself, so frequency was calculated using the best linear fit of the increasing angle. Using this method the correlation coefficient between the experimental data and fit was in most cases in the range between $0.8$ and $0.9$.

Lower precision in the frequency readout was connected also to the fact that the rotation of the puck was not necessarily around the main canonical axis. In many cases the puck was also carrying out precession, which led to deformation of the image of the puck and complications in read-out. However, all these effects were clearly having smaller impact onto the final results than the instability of the shots by individual players.  

\subsection{Measuring friction coefficient}
We have measured the friction coefficients of both Renfrew and Specter directly with a puck of the same type as used during the experiment. We have glued Specter and Renfrew tape on the table on a path of about $25cm$, where for Renfrew we have produced three different surfaces. In one case we simply glued a few long strips along each other (this way it is usually not used during playing, but the result shows basic physical properties of the tape). Two others had short strips glued perpendicular to the motion and partly covering each other, as players do on their blades. Here one can define two different direction of motion, "uphill" and "downhill", where the first corresponds to the way the tape was glued on the blade during our experiments.
We have then slid the puck on the tape, fixed in a special cage holding the puck (of total mass including puck $m_p$) on its side while sliding to prevent its rotation. The cage was fixed with a string via a pulley to a mass $m_w$ and the whole system was let to move freely. The motion was captured by camera and tracked to obtained the acceleration $a$ of the system. The friction was then calculated by 
\begin{equation}
f=\frac{m_pg-(m_p+m_w)a}{m_w g}.
\end{equation}
The results show $f=0.69\pm 0.02$ for Specter, $f=0.46\pm0.07$ for Renfrew longitudinally, $f=0.43\pm0.08$  transversely "downstairs" and  $f=0.50\pm0.11$ transversely "upstairs". Thus we can conclude that the friction coefficient of Specter is significantly higher then of Renfrew.  

\subsection{Statistical Methods} \label{sec:stat}
For each set of about $10$ shots of the same type of the shot and taping method we first defined outliers independently for speed and frequency of rotation, according to (\ref{eq_fences}). Then all shots where either speed or frequency was specified as outlining were removed. This, in most of the cases, led to removal of no more $20\%$ of shots. 

To test the assumptions of t-test, we performed Shapiro-Wilk normality test on speed, rotation and energy gains. Results (Table~\ref{tab:Pvals}) show that normality of all parameters can not be ruled out for Wrist shots ($P>0.05$), unlike for Slap shots, where the data seem to deviate significantly from normality ($P<0.05$).  
Therefore, for Wrist shot gains we further use parametric Student's t-test whereas for Slap shot gains we use non-parametric Wilcoxon (Mann-Whitney) signed rank test \cite{Wasserman2010}. The research hypothesis is that gain in speed, rotation frequency and energy is significantly higher than zero, null hypotheses is that the gain is not positive. 

P-values of statistical tests performed on gains are summarised in Table~\ref{tab:Pvals}. All P-values are much smaller than $0.05$, ruling out the null hypotheses that Specter does not improve performance.

\begin{table}[!ht]
\centering
\begin{tabular}{|l|c|c|c|c|}
\hline
         & test & \multicolumn{1}{c|}{\begin{tabular}[c]{@{}c@{}} Speed \end{tabular}} &
         \multicolumn{1}{c|}{\begin{tabular}[c]{@{}c@{}} Rotation \end{tabular}} &\multicolumn{1}{c|}{\begin{tabular}[c]{@{}c@{}}Energy \end{tabular}} \\ \hline

Wrist Shot &    Shapiro-Wilk &     $0.1363$              &         $0.756$      &       $0.1312$  \\ \hline
Slap Shot &     Shapiro-Wilk &    $0.0067$               &       $0.043$       &       $0.0050$  \\ \hline
Wrist Shot &   t-test &     $0.00032$               &       $0.024$       &       $0.00033$    \\ \hline
Slap Shot &    Wilcoxon &    $0.0012$               &       $0.011$  &       $0.0027$  \\ \hline

\end{tabular}
\caption{ 
\emph{P-values for significance tests.}
First two rows of Table shows results of Shapiro-Wilk normality tests of gains for all parameters for both Wrist and Slap Shots. For Wrist shots, all P-values are greater than $0.05$ meaning the samples may be considered normal. Variances of each pair of samples under consideration can be considered equal according to the variance homogeneity test ($P>0.05$). Thus all Wrist shot samples are considered parametric. However, for Slap Shots, the P-values are smaller than 0.05 suggesting the distributions of gains deviate significantly from normality and therefore Slap shot samples are considered non-parametric.
The last two rows show results of statistical tests of whether gain in Speed, Rotation and Energy is significantly higher than zero. All P-values are smaller than significance level of $0.05$, meaning Specter has a significant effect on gains in all three parameters.}
\label{tab:Pvals}
\end{table}

\subsection*{Data Availability}
The data sets measured and analysed during the current study are available from the corresponding author on reasonable request.

\section*{Author contributions statement and Additional information}
SP and MP conceived and conducted the experiments, SP performed the video analysis and tracking, MP and SP analysed the results, NR and MP performed statistical analysis and designed the draft of the manuscript. All authors contributed to preparation and review of the manuscript.
The authors declare no competing financial interests.

\section*{Acknowledgements}
We would like to express our gratitude to all ice hockey players who voluntarily collaborated on the study and also to Miroslav Lazo, owner of HDC \cite{HDC2018} for lending the facility. MP would like to thank Du\v sana Dokupilov\'a for discussions on statistical analysis of the results obtained. SP would like to thank Martin Marek for borrowing the experimental equipment.


\bibliographystyle{unsrt}
\bibliography{bibliography.bib}

\end{document}